\documentclass[twocolumn]{aastex63}
\usepackage{rotating}
\usepackage{graphicx}
\usepackage{amssymb}
\usepackage{amsmath}
\usepackage{hyperref}
\usepackage{lineno}


\def\gtrsim{\mathrel{\hbox{\rlap{\hbox{\lower4pt\hbox{$\sim$}}}\hbox{$>$}}}}
\def\lesssim{\mathrel{\hbox{\rlap{\hbox{\lower4pt\hbox{$\sim$}}}\hbox{$<$}}}}

\begin{document}

\title{Precise timing and phase-resolved spectroscopy of the young pulsar 
J1617--5055 with {\sl NuSTAR}}
\author{Jeremy Hare}
\altaffiliation{NASA Postdoctoral Program Fellow}
\affiliation{NASA Goddard Space Flight Center, Greenbelt,  MD 20771, USA}
\author{Igor Volkov}	
\affiliation{Department of Physics, The George Washington University, 725 21st St. NW, Washington, DC 20052}
\author{George G. Pavlov}
\affiliation{Department of Astronomy \& Astrophysics, Pennsylvania State University, 525 Davey Lab, University Park, PA 16802, USA}
\author{Oleg Kargaltsev}
\affiliation{Department of Physics, The George Washington University, 725 21st St. NW, Washington, DC 20052}
\affiliation{The George Washington Astronomy, Physics, and Statistics Institute of Sciences (APSIS)}
\author{Simon Johnston}
\affiliation{CSIRO Astronomy \& Space Science, Australia Telescope National Facility, P.O. Box 76, Epping, NSW 1710, Australia}

\email{jeremy.hare@nasa.gov}

\begin{abstract}
We report on a NuSTAR observation of the young, energetic pulsar PSR J1617--5055. Parkes Observatory 3 GHz radio observations of the pulsar (taken about 7 years before the NuSTAR observations) are also reported here. NuSTAR detected pulsations at a frequency of $f\approx14.4$ Hz ($P\approx69.44$ ms) and, in addition, the observation was long enough to measure the source's frequency derivative, $\dot{f}\approx-2.8\times10^{-11}$ Hz s$^{-1}$. We find that the pulsar shows one peak per period at both hard X-ray and radio wavelengths, but that the hard X-ray 
pulse is broader (having a duty cycle of $\sim 0.7$), than the radio pulse (having a duty cycle of $\sim 0.08$). Additionally, the radio pulse is strongly linearly polarized. J1617's phase-integrated hard X-ray spectrum is well fit by an absorbed power-law model, with a photon index $\Gamma=1.59\pm 0.02$. The hard X-ray pulsations are well described by three Fourier harmonics, and have a pulsed fraction that increases with energy. We also fit the phase-resolved NuSTAR spectra with an absorbed power-law model in five phase bins and find that the photon index varies with phase from $\Gamma = 1.52\pm 0.03$ at phases around the flux maximum to  $\Gamma=1.79\pm 0.06$ around the flux minimum. Lastly, we compare our results with other pulsars whose magnetospheric emission is detected at hard X-ray energies and find that, similar to previous studies, J1617's hard X-ray properties are more similar to the MeV pulsars than the GeV pulsars.  
\end{abstract}

\section{Introduction}

Broadband X-ray observations of young, energetic pulsars provide a means to probe  particle acceleration in pulsar magnetospheres. Previous studies, using phase-resolved spectroscopy, have shown that  the  
spectral slopes of young pulsars' X-ray spectra vary as a function of both phase and energy (see e.g., \citealt{2000A&A...361..695M,2003A&A...400.1013D,2006A&A...450..617M}). In particular, hard X-ray observations
help to establish the spectral link between the soft X-ray and $\gamma$-ray energy range (see e.g., Figure 3 in \citealt{2020MNRAS.492.1025C}). Increasing the sample size of pulsars studied at these energies allows one to understand how the pulsar's magnetospheric emission depends on the  pulsar's parameters,
such as age, spin-down power, magnetic field, and the orientation of the spin and magnetic axes (see e.g., \citealt{2020MNRAS.492.1025C}).  

In the past, studies of these sources at hard X-ray energies have been hampered  by the high background of  non-imaging instruments, making it difficult to measure  the spectrum of their off-pulsed emission,  especially when there is contamination from a bright pulsar wind nebula (see e.g., \citealt{2015MNRAS.449.3827K}). However, the improved angular resolution and sensitivity afforded by the {\sl Nuclear Spectroscopic Telescope Array} ({\sl NuSTAR}), has made it possible to study  these pulsars in exquisite detail, often even allowing for the pulsar wind nebulae (PWNe) surrounding these pulsars to be resolved (\citealt{2014ApJ...789...72N,2014ApJ...793...90A,2015ApJ...801...66M,2016ApJ...817...93C}). 

Pulsations from PSR J1617--5055 (J1617 hereafter) with a period $P\approx 69$ ms
were first 
discovered 
in X-rays with
{\sl Ginga} and {\sl ASCA}  \citep{1992IAUC.5588....2A,1998ApJ...494L.207T},
and then in the radio with the Parkes Observatory
\citep{1998ApJ...503L.161K}.
It is a young ($\tau=8.3$ kyr), energetic ($\dot{E}=1.6\times10^{37}$ erg s$^{-1}$) pulsar with 
a magnetic field $B=3.1\times10^{12}$ G. Based on its dispersion measure ${\rm DM}\simeq467$ pc\,cm$^{-3}$, the distance to the pulsar was
estimated to be 4.7 kpc 
or 6.8 kpc for the Galactic free electron-density distribution models by \cite{2017ApJ...835...29Y} or \cite{2002astro.ph..7156C}, respectively.

X-ray observations of J1617 at different energies have shown
evidence of spectral softening  of  the phase-integrated spectrum with  energy:
the photon index varied from $\Gamma\approx1.1$ in the 2--10 keV band to $\Gamma\approx1.9$ in the 20--300 keV band \citep{2007MNRAS.380..926L,2009ApJ...690..891K,2015MNRAS.449.3827K}. Pulsed X-ray emission has been detected  from J1617 with a large pulsed  fraction of $\sim80\%$ at  
lower (2--10 keV) energies, while  at 
higher energies pulsations were  marginally detected up  to 
$\gtrsim30$ keV \citep{2002nsps.conf...64B,2015MNRAS.449.3827K}.
J1617 is also  notable for  exhibiting  large glitches \citep{2000ApJ...534L..71T} and having a relatively under-luminous X-ray PWN  \citep{2009ApJ...690..891K}.

Although J1617 is 
 young, no related X-ray or radio supernova remnant  has been detected in its vicinity. Pulsed emission from
J1617 has not  been detected at GeV energies  to date\footnote{See \url{https://confluence.slac.stanford.edu/display/GLAMCOG/Public+List+of+LAT-Detected+Gamma-Ray+Pulsars}}. Additionally, despite its low luminosity in X-rays
\citep{2009ApJ...690..891K}, 
  several authors have suggested that J1617's PWN  contributes, at least partially, to the emission of the extended TeV source HESS  J1616--508 (see \citealt{2006ApJ...636..777A,2007MNRAS.380..926L,2017ApJ...841...81H}). However, the large spatial offset ($\sim10'$) and asymmetry of the  X-ray PWN shed some doubt  on this association. 

In this paper, we report on the results of  a {\sl NuSTAR} observation of J1617, and the results of the most recent radio timing campaign carried out by Parkes, but preceding the NuSTAR observation. In Section \ref{obs_and_dat} we describe  the observation and data reduction, while in Section \ref{results} we report on the X-ray and radio timing and X-ray spectral analysis. In Section \ref{discuss}, we discuss our results in the context of hard X-ray observations of other young, energetic pulsars. A description of our approach to the 
 X-ray timing  is provided in the Appendix.

\section{Observations and Data Reduction}
\label{obs_and_dat}

\subsection{NuSTAR}

NuSTAR (\citealt{2013ApJ...770..103H}) 
observed J1617 from 2018 April 29 to 2018 May 2 (starting at MJD 58237.69, obsID 30301013002) for $\approx131$ ks of scientific exposure time (PI G. Pavlov). The data were reduced using 	the {\sl NuSTAR} Data Analysis Software (NuSTARDAS) package version 2.0.0 and the 20200912 version of the Calibration database (CALDB). Prior to performing 
data analysis, we corrected the photon arrival times to the solar system barycenter using {\tt nupipeline}. This tool also corrects for {\sl NuSTAR's} clock drift\footnote{See \url{http://www.srl.caltech.edu/NuSTAR_Public/NuSTAROperationSite/clockfile.php}} and provides a clock accuracy of about 65 $\mu$s \citep{2020arXiv200910347B}. The source's energy spectra and event lists were extracted from both of the FPMA and FPMB detectors using a $r=72''$ circle centered on the source.  This radius was chosen to include as many source counts as possible, while also avoiding stray-light that fell on the same chip as the source in the FPMA detector. The corresponding background energy spectra and event lists were extracted from a source-free circular region ($r=90''$) on the same detector chip as J1617. 

We note here that we attempted to include soft X-ray data from {\sl XMM-Newton} and {\sl Chandra} in our phase-integrated spectral fits. However, the photon indices derived from independent fits (i.e., not jointly fit) to the {\sl XMM-Newton} and {\sl NuSTAR} spectra in the overlapping 3--10 keV band, and the 3--8 keV {\sl Chandra} spectrum, disagreed by $\sim2\sigma$--$3\sigma$, leading to systematic residuals in the joint fits  (primarily at lower energies). This discrepancy is likely due to calibration uncertainties between the observatories, so we exclude the soft X-ray data. In any case, since the source is highly absorbed (see Section \ref{xspectro}), the soft X-ray data only extend the energy coverage down to $\sim 1$ keV. Therefore, 
including the soft X-ray data 
in our analysis 
would not significantly impact the derived spectral parameters.

\subsection{Radio Observations }
Observations of the pulsar were made with the Parkes radio telescope under the auspices of project P574. Data collection started in early 2007 and continued until early 2014, observations took place approximately monthly during that period. The majority of the observations were carried out at a central frequency of 1369~MHz with 256~MHz of bandwidth. At this frequency the pulsar suffers from scatter-broadening of its profile due to the interstellar medium. Sporadically, therefore, supplementary observations were made at a frequency of 3096~MHz with a bandwidth of 1024~MHz. Details of the data reduction and calibration process can be found in \citet{2010PASA...27...64W} and \citet{2018MNRAS.474.4629J}.

\section{Results}
\label{results}

\subsection{X-ray Timing}
\label{X-ray timing sec}

The time of arrival for the first event
detected by {\sl NuSTAR} was MJD 58237.70488341 TDB 
(or $t_0=262716835.741295$ s in {\sl NuSTAR} time), which we define as  the start of the observation.
While the scientific exposure time was 131 ks, the total length spanned by the observation, defined as the time elapsed between the first and last detected event, was $T_{\rm span}=246541.2961$ s or about 2.85 days. 
 The large interval spanned by the observation has allowed us 
 not only to measure
 the frequency $f$ 
 with high precision but also to measure the
 frequency derivative $\dot{f}$. To minimize the correlation between $f$ and $\dot{f}$,
we subtract ($t_0+T_{\rm span}/2$) from the event times 
prior to the $f$, $\dot{f}$ search, so that the measured 
ephemeris corresponds to the temporal center of the data set (i.e., MJD 58239.13162703 TDB).

\begin{figure}
\centering
\includegraphics[trim={0 0 0 0},scale=0.45]{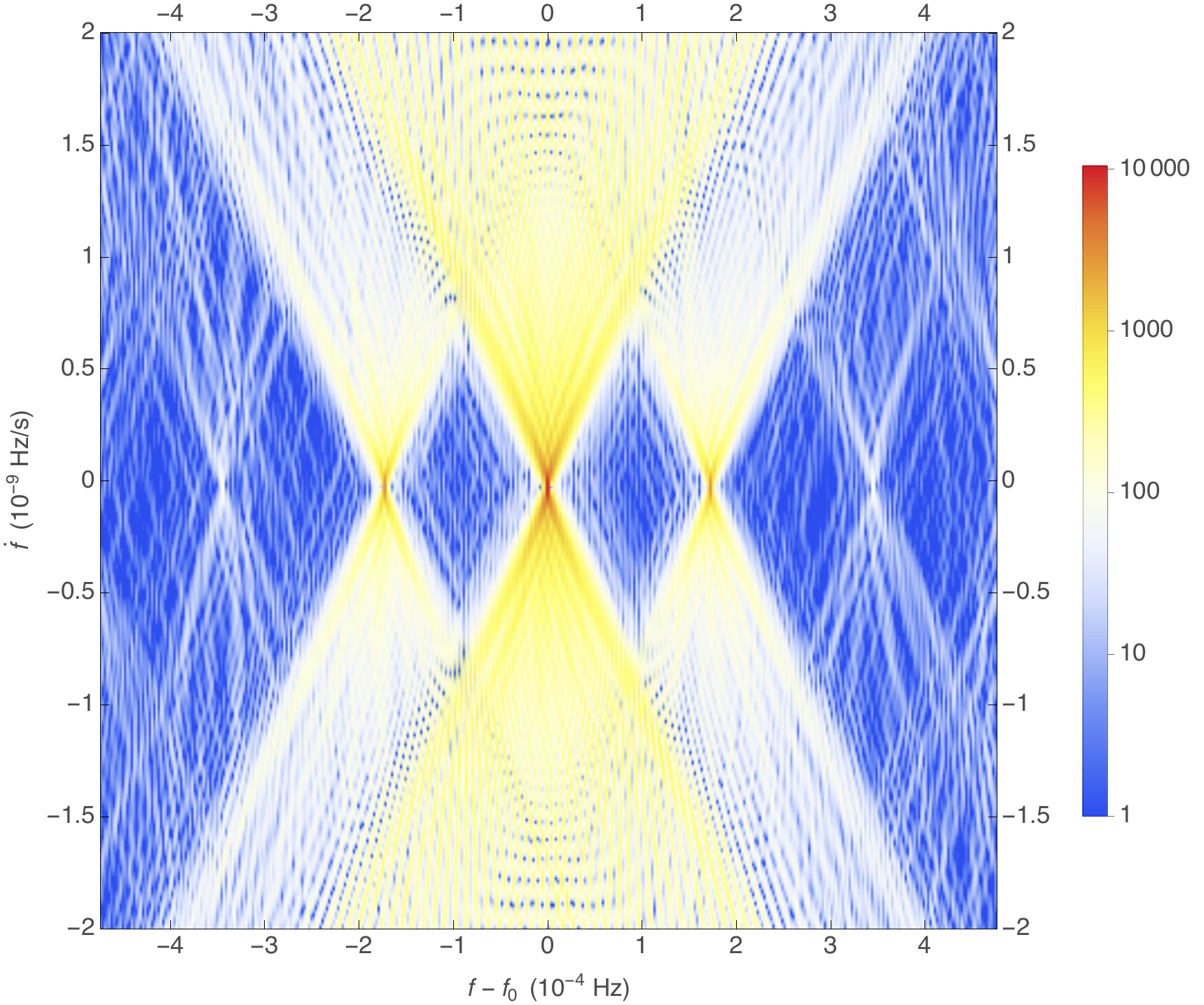}
\centerline{\includegraphics[trim={0 0 0 0},scale=0.45]{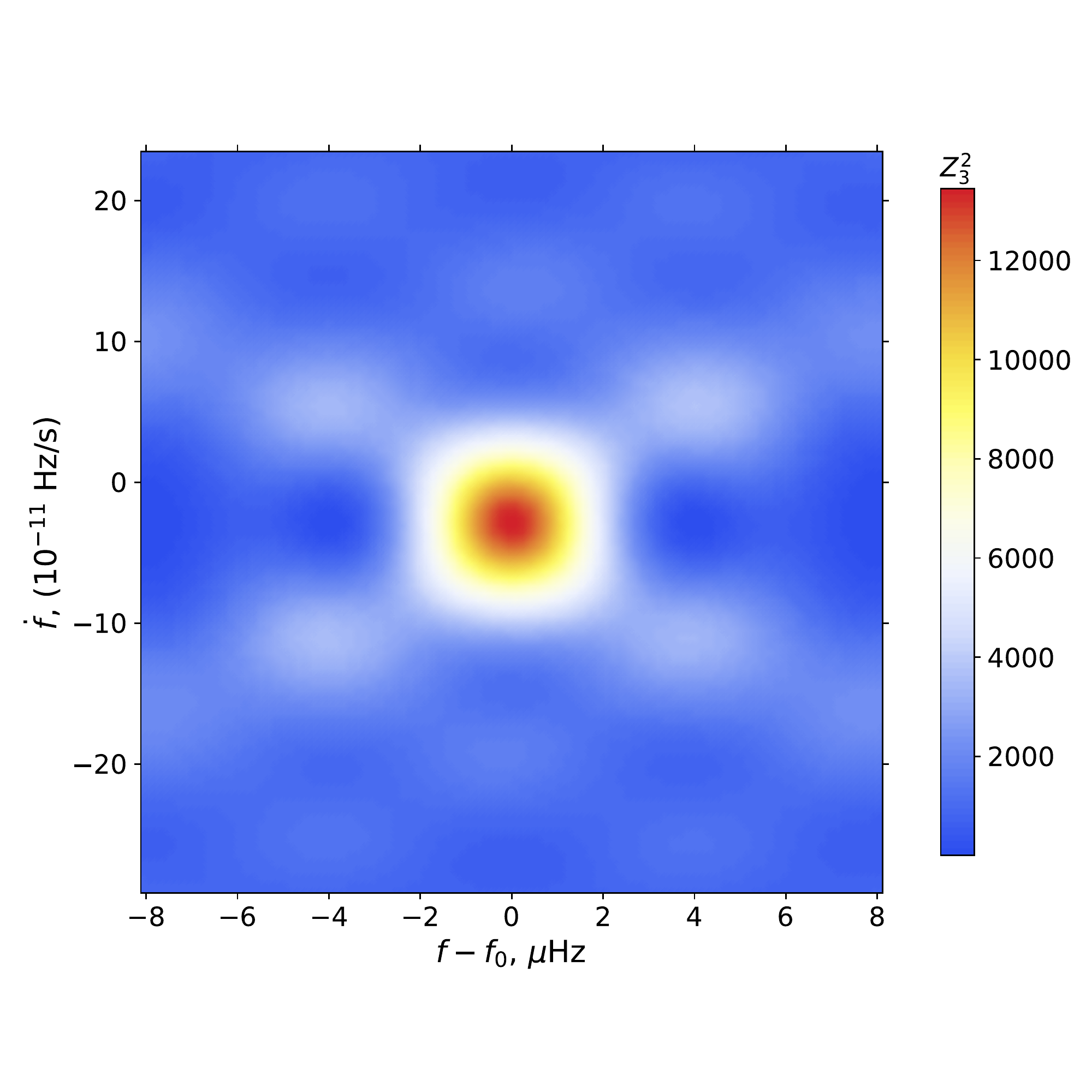}}
\caption{Zoomed out (top) and zoomed in (bottom) maps of the $Z^{2}_3(f,\dot{f})$ surface in the vicinity of the pulsar's frequency and its derivative,
for events in the 3--40 keV energy range. 
The centers of the maps, corresponding to the top of the main peak, 
$Z_{3,\rm max}^2 = 13495$, are at the $f, \dot{f}$ given by Equation (\ref{ephem_sol}).  The sizes of the base of the peak along the $f$ and $\dot{f}$ directions are determined by $T_{\rm span}^{-1}\approx 4.1$ $\mu$s and $2 T_{\rm span}^{-2} \approx 3.3\times 10^{-11}$ Hz/s, respectively.
\label{fig:f_fdot_z}
}
\end{figure}

Based on previous observations, J1617's spin frequency and its derivative are
anticipated to be about 14.400 Hz and $-2.8\times 10^{-11}$ Hz s$^{-1}$ at the {\sl NuSTAR} observation epoch.
To find the best-fit 
values and uncertainties of $f$ and $\dot{f}$, 
we used the $Z^{2}_k$ test \citep{1983A&A...128..245B}, where $k$ is the number of harmonics 
included in the test (see the Appendix for details).
We calculated the $Z_k^2$ values on a grid in the $f$-$\dot{f}$ plane in the vicinity of their expected values for {\sl NuSTAR} events in different energy bands, and for different numbers of  harmonics  (see Table \ref{tab:z_3_v_ene}).
We found that the maximal $Z_k^2$ values are obtained using events in the 3--40 keV energy range, which  provides the most precise measurement of $f$ and $\dot{f}$.
We then used the H-test 
(see, e.g., \citealt{1989A&A...221..180D}) and found that the H-statistic reaches a maximum for $k=3$. 
 As a result, we found
 \begin{equation}
 \label{ephem_sol}
     f_0=14.40012559(2)\, {\rm Hz},\quad \dot{f_0}=-2.83(6)\times10^{-11}\,{\rm Hz}\,{\rm s}^{-1},
 \end{equation}
with a corresponding $Z_{\rm 3,max}^2 = 13495$. The $1\sigma$ uncertainties  were determined through Monte-Carlo (MC) simulations (see the Appendix for  details). Throughout this paper, the last digit in parentheses corresponds to the $1 \sigma$ uncertainty on  the last digit of the measured quantity. 
Maps of the $Z_3^2(f,\dot{f})$ surface in the vicinity of $f_0$ and $\dot{f}_0$ are shown in Figure \ref{fig:f_fdot_z} on two different scales. The alias peaks and ridges in the zoomed out map are caused by the multiple gaps due to Earth occultations during the {\sl NuSTAR} observation.
Notice that, due to the large height of the $Z_3^2$ peak,
the $f$ and $\dot{f}$ uncertainties
are so small that they cannot be resolved even in the zoomed in map 
(see  Equation \ref{eq:deltaffdot}).

\begin{figure}
\centering
\includegraphics[trim={50 0 0 0},scale=0.5]{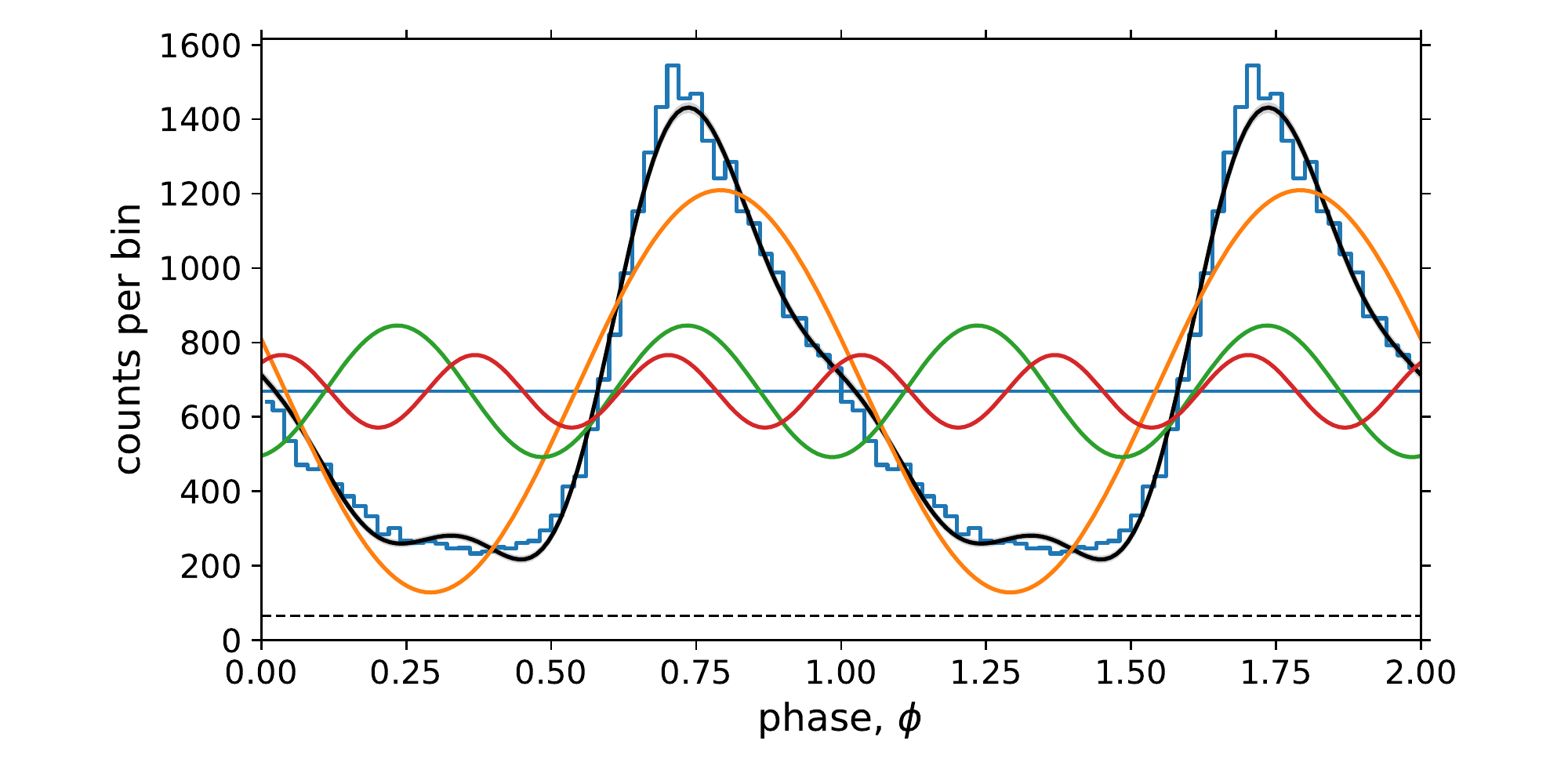}
\caption{ 
X-ray pulse profile (phase-folded light curve) for J1617 in the 3--40 keV band plotted as a blue histogram with 50 phase bins and as
the sum of 3 harmonics, $(N/50){\cal F}(\phi)$ (black curve; see Equation (\ref{eq:fourier}) in the Appendix), where $N=33448$ is the total number of events.
The first, second, and third harmonics are shown in orange, green, and red, respectively. 
The background 
of 65.9 counts per bin is shown by a dashed line.
\label{fig:pulse_profile}
}
\end{figure}

The phase-folded light curve in the 3--40 keV energy band and the contributions from each harmonic are shown in Figure \ref{fig:pulse_profile}. 
The comparison of the 50-bin histogram with the sum of 3 harmonics (plus a DC level) is in 
good agreement between these two descriptions of the pulse profile, as expected. 
This implies that binning is not required to investigate the pulse profile in such a case of smooth pulsations.
Using Fourier analysis without binning, which is particularly convenient when only a few harmonics are important, allows one to avoid the additional uncertainties associated with the binning procedure.

\begin{figure}[h] 
\centering
\includegraphics[trim={50 0 0 0},scale=0.5]{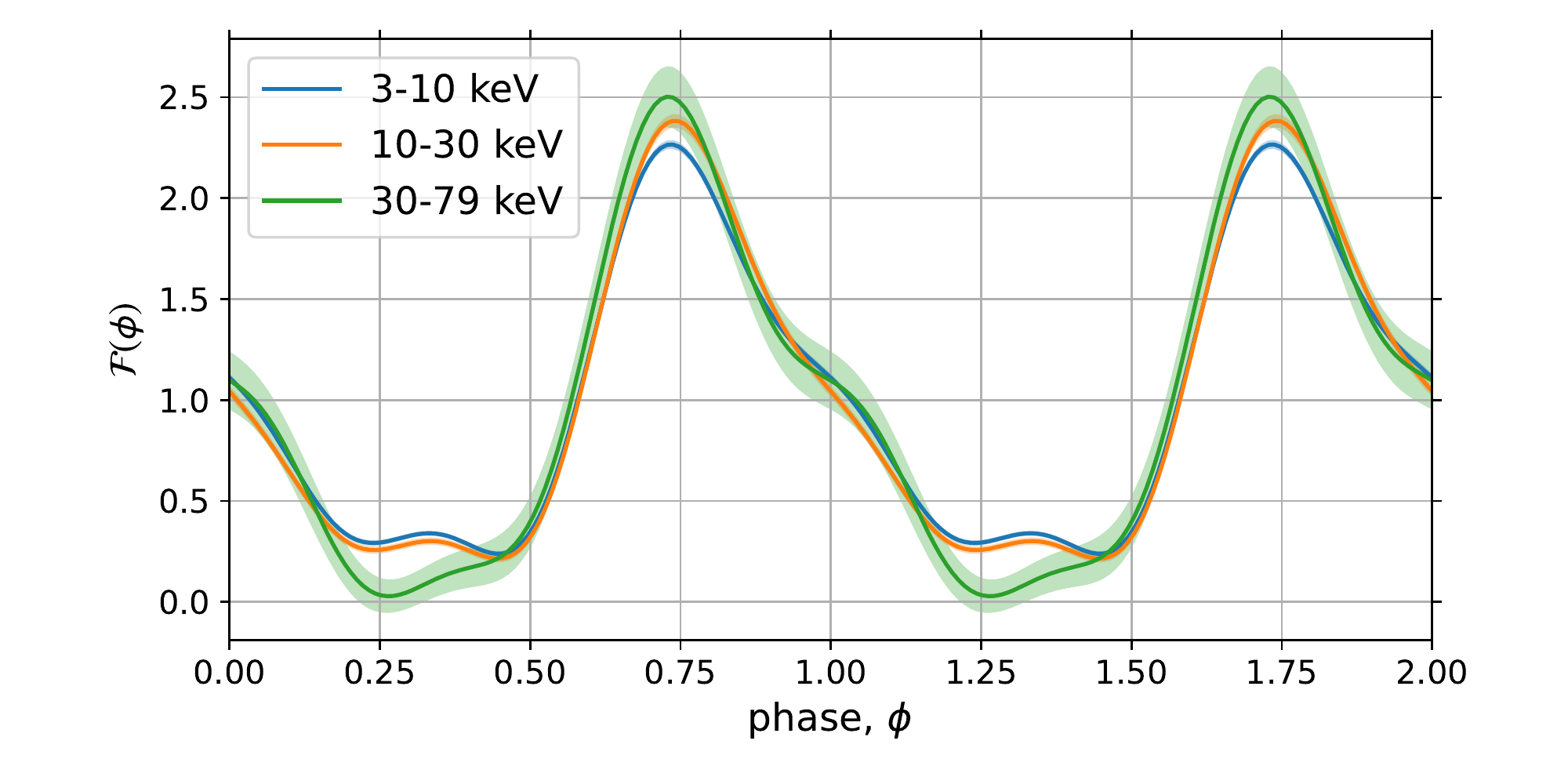}
\caption{Normalized pulse profiles ${\cal F}(\phi)$ for several energy bands. All plots are corrected for background. Shaded areas correspond to $\pm1$ standard  deviation for the ${\cal F}$ at given $\phi$ (found from MC simulations).
}
\label{fig:all_ampls_corr}
\end{figure}

Figure \ref{fig:all_ampls_corr} shows the normalized pulse profiles ${\cal F}(\phi)$ (see Equation \ref{eq:fourier} in the Appendix for the definition) and their standard deviations in  3 energy bands. We see that the shapes of the light curves are very similar to each other, but their  amplitudes (hence pulsed fractions) slightly grow with energy.

\begin{figure}
\centerline{\includegraphics[trim={0 0 0 0},scale=0.6]{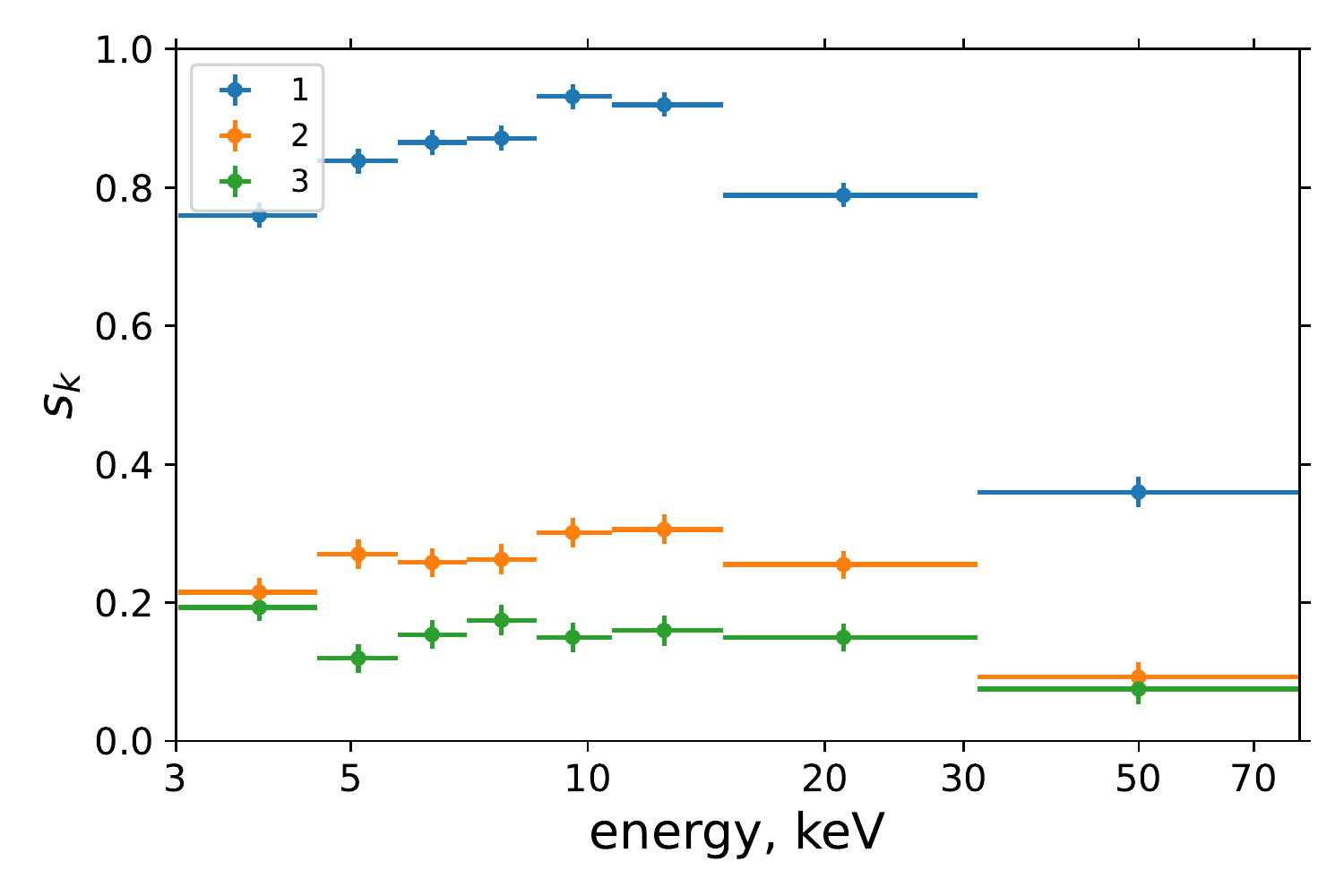}}
\centerline{\includegraphics[trim={0 0 0 0},scale=0.6]{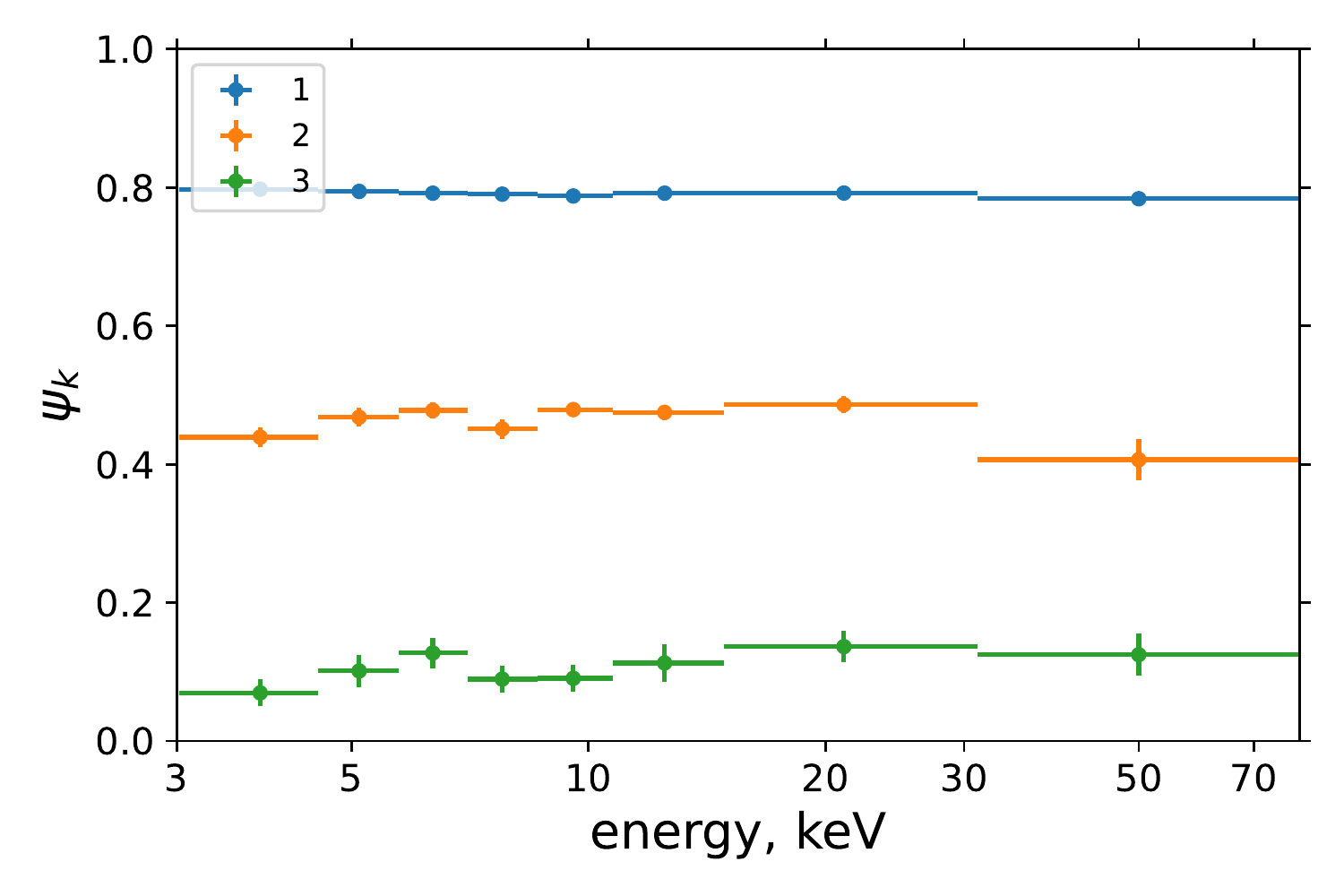}}
\caption{
The energy dependence for Fourier amplitudes $s_k$ 
and phases $\psi_k$ of 3 signal harmonics, without correction for background.  
\label{fig:sk_and_phase}
}
\end{figure}

Figure \ref{fig:sk_and_phase} shows the energy dependencies of the amplitudes $s_k$ and phases $\psi_k$ of the three harmonics.
We see that the amplitude $s_1$, which is considerably larger than $s_2$ and $s_3$, grows with energy at $E\lesssim 10$ keV; its decline at higher energies is caused by relative increase of background (which is uncorrected for). The phases $\psi_k$ of the 3 harmonics determine the harmonic maxima, $\phi_{k, \rm max}^j = (\psi_k + j)/k$, where $j$ is an integer number.  Since they do not show an appreciable dependence on energy for J1617, we can measure them in a larger energy range with a higher precision. For instance,
for the 3--40 keV band, we obtained
$\psi_1=0.7913\pm0.0013$,
$\psi_2=0.4688\pm 0.0045$,
and $\psi_3=0.1043\pm 0.0082$.

An important property of a folded light curve is the pulsed fraction, which, generally, may depend on the harmonic amplitudes and phases, and their dependence on energy.
We 
calculated three  commonly used 
``pulsed  fractions'', namely, the area pulsed fraction $p_{\rm area}$, the light curve amplitude $p_{\rm amp}$, and the root-mean-square (rms) deviation of the pulsating flux from its DC (mean) value that is often called  ``the rms pulsed fraction'' $p_{\rm rms}$
(see the Appendix C for their definitions). 
The pulsed fraction calculated from 
timing of events extracted from the source aperture 
is generally lower than the intrinsic one because of the (unpulsed) background contribution.
To correct for the background contribution, 
we multiply the measured pulsed fractions by the ratio
$N/N_s$, where $N$ is the measured (source + background)  number of counts in the source aperture,  $N_s = N - N_b$ is the number of source counts,
and $N_b$ is the number of background counts scaled to the source aperture area.

\begin{figure}
\centerline{\includegraphics[trim={0 0 0 0},scale=0.6]{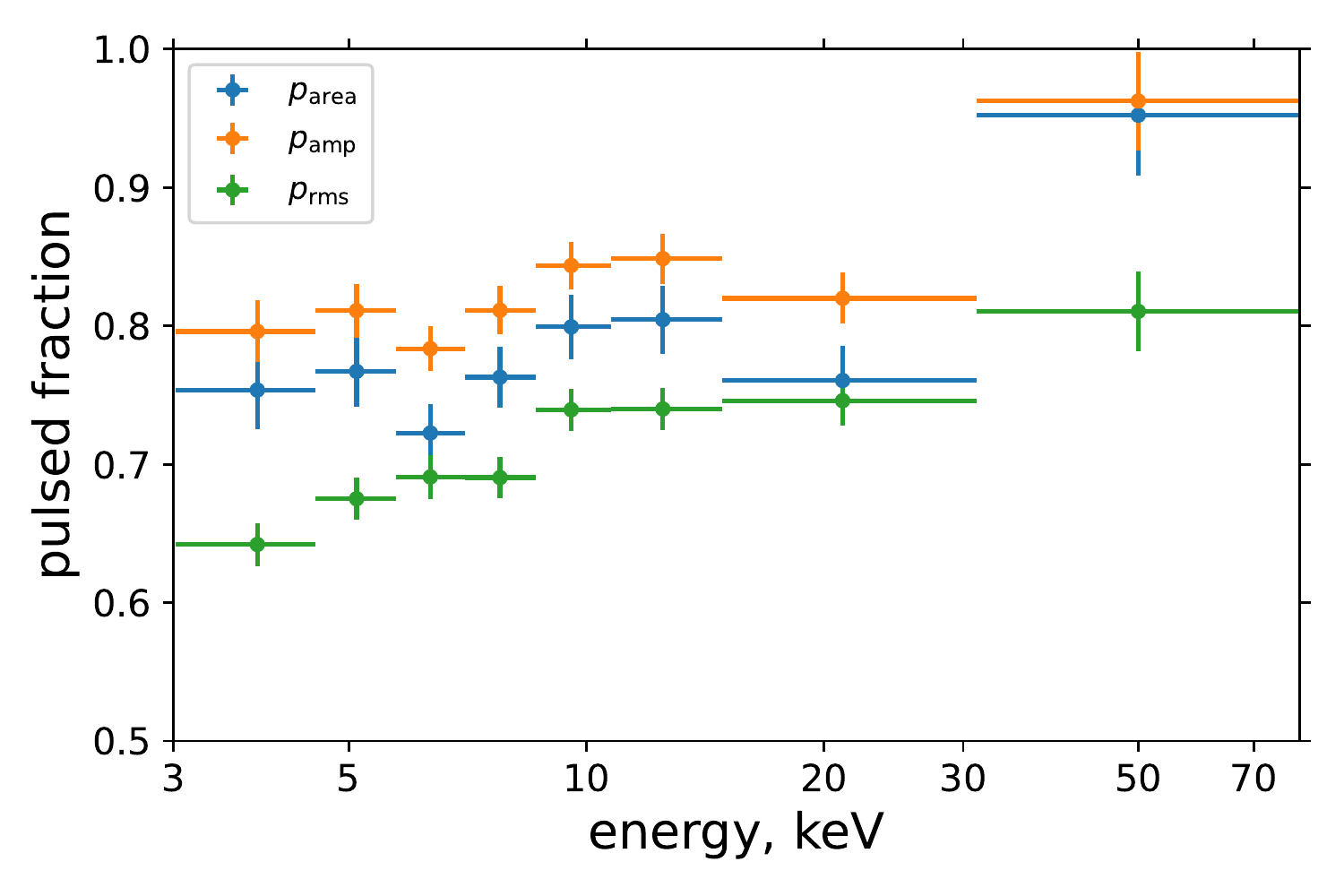}}
\caption{Energy dependence of three background-corrected (intrinsic) pulsed fractions 
(see the Appendix C for  definitions).}
\label{fig:PF_rms}
\end{figure}

Table \ref{tab:PF_v_ene} provides the  observed pulsed fraction for each pulsed fraction definition, as well as the background correction factors ($N/N_s$) in several energy bands. The intrinsic pulsed fractions are shown as a function of energy in Figure \ref{fig:PF_rms}.
Regardless of  the
adopted pulsed fraction definition, 
the  intrinsic pulsed fraction  increases with energy.

\begin{deluxetable}{lccccc}
\tablecolumns{6}
\tablecaption{Maximum values of $Z^{2}_k(f,\dot{f})$ 
for different energy ranges and numbers of  harmonics ($k$)
} 
\label{tab:z_3_v_ene}
\tablewidth{0pt}
\tablehead{
\colhead{Energies} & \colhead{$N$} & 
\colhead{$Z^{2}_1$}  & \colhead{$Z^{2}_2$} & 
\colhead{$Z^{2}_3$} & \colhead{$Z_{4}^2$}  \\
\colhead{ keV} & \colhead{counts} & \colhead{} & \colhead{}& \colhead{} & \colhead{} 
}
\startdata
 3--10 & 21663& 7859 & 8721& 8770 &  8772\\
 10--30 & 10738 & 4063 & 4555 & 4579 & 4583\\
 30--79 & 2704 &179 & 194 & 194 & 197\\
 3--79 & 35105 & 11796 & 13113 & 13179 & 13181\\
 \hline
 3--40 & 33448 & 12062  & 13423 & 13495 & 13497\\
\enddata
\end{deluxetable}

\begin{deluxetable}{lcccc}
\tablecolumns{5}
\tablecaption{ Observed pulsed fraction calculated using various definitions (see Appendix C) and background correction factors (see Section \ref{X-ray timing sec}) in 4 energy bands.
} 
\label{tab:PF_v_ene}
\tablewidth{0pt}
\tablehead{
\colhead{} & \colhead{3--10 keV} & \colhead{10--30  keV}  & \colhead{30--79 keV} &     \colhead{3--40  keV}
}
\startdata
 $p_{\rm area}$ & 0.71(1) & 0.69(2)  & 0.31(4) & 0.69(1)\\
 $p_{\rm amp}$ & 0.75(1) & 0.73(1) & 0.32(3) & 0.74(1) \\
 $p_{\rm rms}$ & 0.64(1) & 0.65(1) & 0.27(5) & 0.63(1)\\
 \tableline
$N /N_{s}$ & 1.075 & 1.134 & 3.030 & 1.110 \\
 \enddata
 \tablecomments{ The intrinsic (i.e., background corrected) pulsed fraction can be obtained by multiplying the observed pulsed fraction by the background correction factor (see Section \ref{X-ray timing sec} and Figure \ref{fig:PF_rms}).}
\end{deluxetable}

\subsection{Radio Timing}
Radio pulsations from J1617 were first detected by \cite{1998ApJ...503L.161K} using the Parkes telescope, but only a crude timing ephemeris was provided in that paper. More recently, the Parkes telescope timed J1617 between 2007 April 30 and 2014 February 20. During this $\sim7$ year observing period 5 glitches were detected, suggesting that the pulsar glitches roughly once a year \citep{2021MNRAS.tmp.2433L}. The glitches range in strength with the smallest having $\Delta f/f\approx5\times10^{-10}$ and the largest having $\Delta f/f=2\times10^{-6}$, which falls within the typical range of pulsar glitch sizes (see e.g., \citealt{2017A&A...608A.131F}). In addition to the glitches, the pulsar has a very high level of stochastic wander in the times-of-arrival from the pulsar (a phenomenon known as `timing noise'). Both the glitches and the timing noise make constructing an ephemeris difficult, and means that extrapolation beyond the end of the data span extremely problematic. Unfortunately therefore, the radio ephemeris from these (and earlier) observations cannot be used to predict the pulsar frequency or frequency derivative at the epoch of our NuSTAR observation with confidence and prevents us from phase-connecting the radio and X-ray pulse profiles. 

\begin{figure}
\centering
\includegraphics[trim={0 0 0 0},scale=0.45]{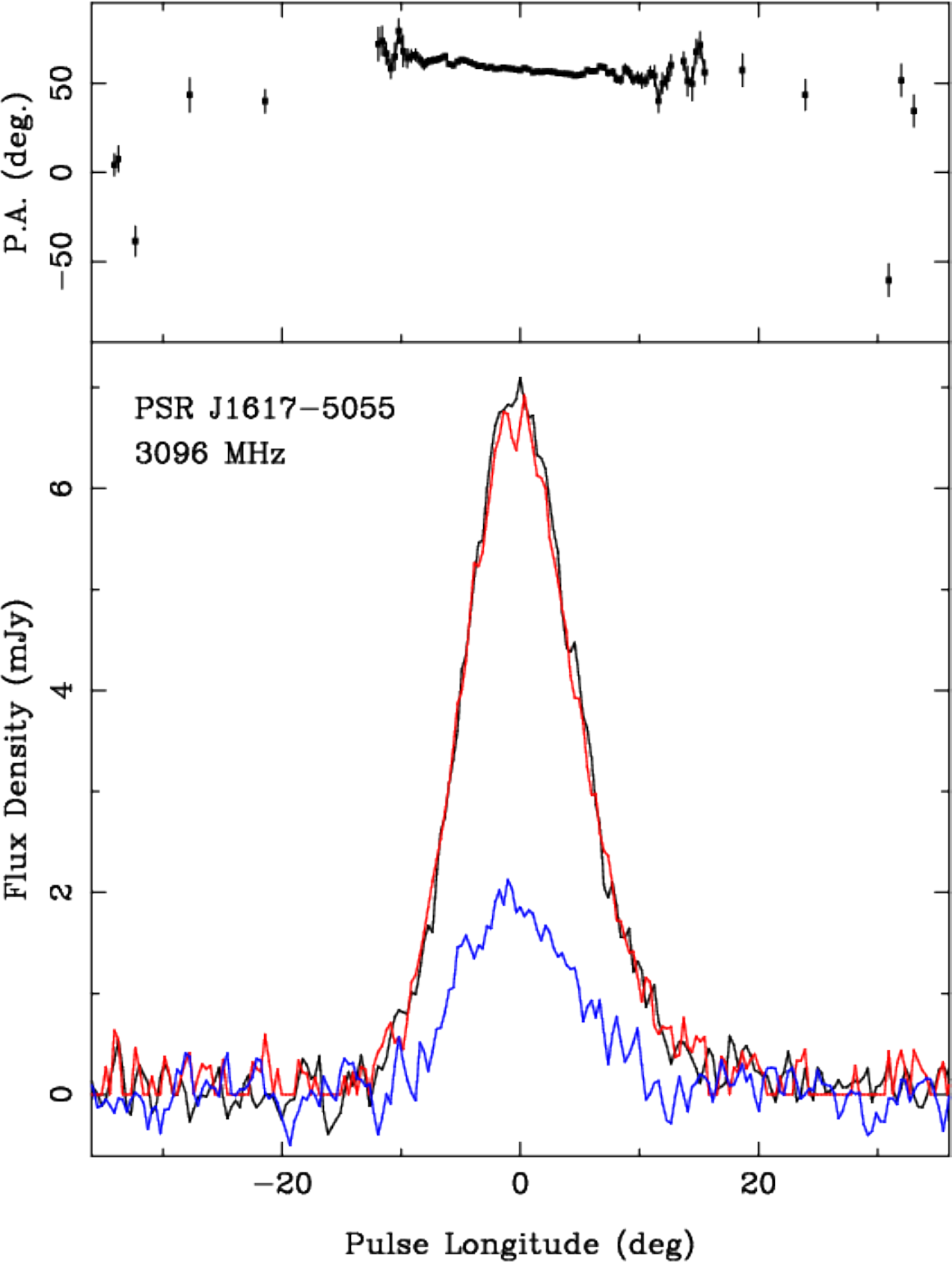}
\caption{{\sl Bottom:} 
Pulse profile of J1617 obtained with the Parkes radio telescope at an observing frequency of 3096~MHz. The black, red, and blue curves show the total intensity, linear and circular polarized components respectively. One pulsar rotation corresponds to $360^\circ$ change of pulse longitude.
{\sl Top:} Position angle of the linear polarization as a function of pulse longitude.
\label{rad_pulse}
}
\end{figure}

The pulse profile in total intensity, as well as the linearly and circularly polarized components from Parkes data taken at 3096~MHz are shown in Figure \ref{rad_pulse}. At this frequency, the profile consists of a single narrow peak with a FWHM of $\approx  0.03 P$. The radio profile is almost 100\% linearly polarized. The high polarization fraction is typical of pulsars with high $\dot{E}$ \citep{2018MNRAS.474.4629J}, though the single-component, narrow profile is different from most other pulsars of its type \citep{2006MNRAS.368.1856J}.
The phase dependence of the position angle of linear polarization makes only a shallow traverse. This may imply that the inclination of the magnetic and rotation axes ($\alpha$) is small and/or the line-of-sight cuts through the edge of the beam. A low value of $\alpha$ may help explain the lack of $\gamma$-ray emission \citep{2020MNRAS.497.1957J}.

\subsection{X-ray spectroscopy}
\label{xspectro}
Prior to fitting, J1617's energy spectra were grouped to have a signal to noise ratio of at least 5 per energy bin. All X-ray energy spectra were fit using XSPEC version 12.11.1 \citep{1996ASPC..101...17A}. The interstellar absorption 
was accounted for with the T\"ubingen-Boulder ISM absorption model ({\tt tbabs}) using solar abundances adopted from \cite{2000ApJ...542..914W}. In all fits we included a constant in the model to account for calibration differences between the FPMA and FPMB detectors and found that this value remains 
$<8\%$ throughout all of our fits. 

\subsubsection{Phase-integrated spectroscopy}
\label{phase_int_spec}
We fit the phase-integrated spectrum of J1617 with an absorbed power-law model. The best-fit model ($\chi^2/\nu=596.4/589 = 1.01$; where $\nu$ is the number of degrees of freedom) has an absorption $N_{\rm H}=6.5(7)\times10^{22}$ cm$^{-2}$ and photon index $\Gamma=1.59(2)$. The unabsorbed flux of the source is $F_{\rm 3-79 \ keV}=1.61(3)\times10^{-11}$ erg cm$^{-2}$ s$^{-1}$. This photon-index is roughly consistent with those measured previously at hard X-ray energies by non-imaging instruments (see, e.g., \citealt{2007MNRAS.380..926L}).

It is important to note that {\sl NuSTAR} does not have the angular resolution necessary to resolve the pulsar emission from the compact PWN emission; therefore, this spectrum consists of both emission components. In the {\sl Chandra} observations of the source,  \cite{2009ApJ...690..891K} were able to disentangle the pulsar and PWN emission. They found that the spectra from both the pulsar and PWN could be well fit by an absorbed power-law model, with the pulsar's spectrum being harder than the PWN's spectrum: $\Gamma_{\rm psr}=1.14(6)$ versus $\Gamma_{\rm PWN}\approx1.7(2)$. Extrapolating the absorbed fluxes from \cite{2009ApJ...690..891K} to the 3--79 keV band, we expect that the PWN could contribute up to $\sim10\%$ of the source's flux compared to the pulsar. Hence the PWN spectrum cannot significantly alter the photon index of the phase-integrated spectrum (see Section \ref{phase_res_spec_dis}).

\subsection{Phase-resolved spectroscopy}
\label{pr_spec_sec}

\begin{figure}
\includegraphics[trim={0 0 0 0},scale=0.31]{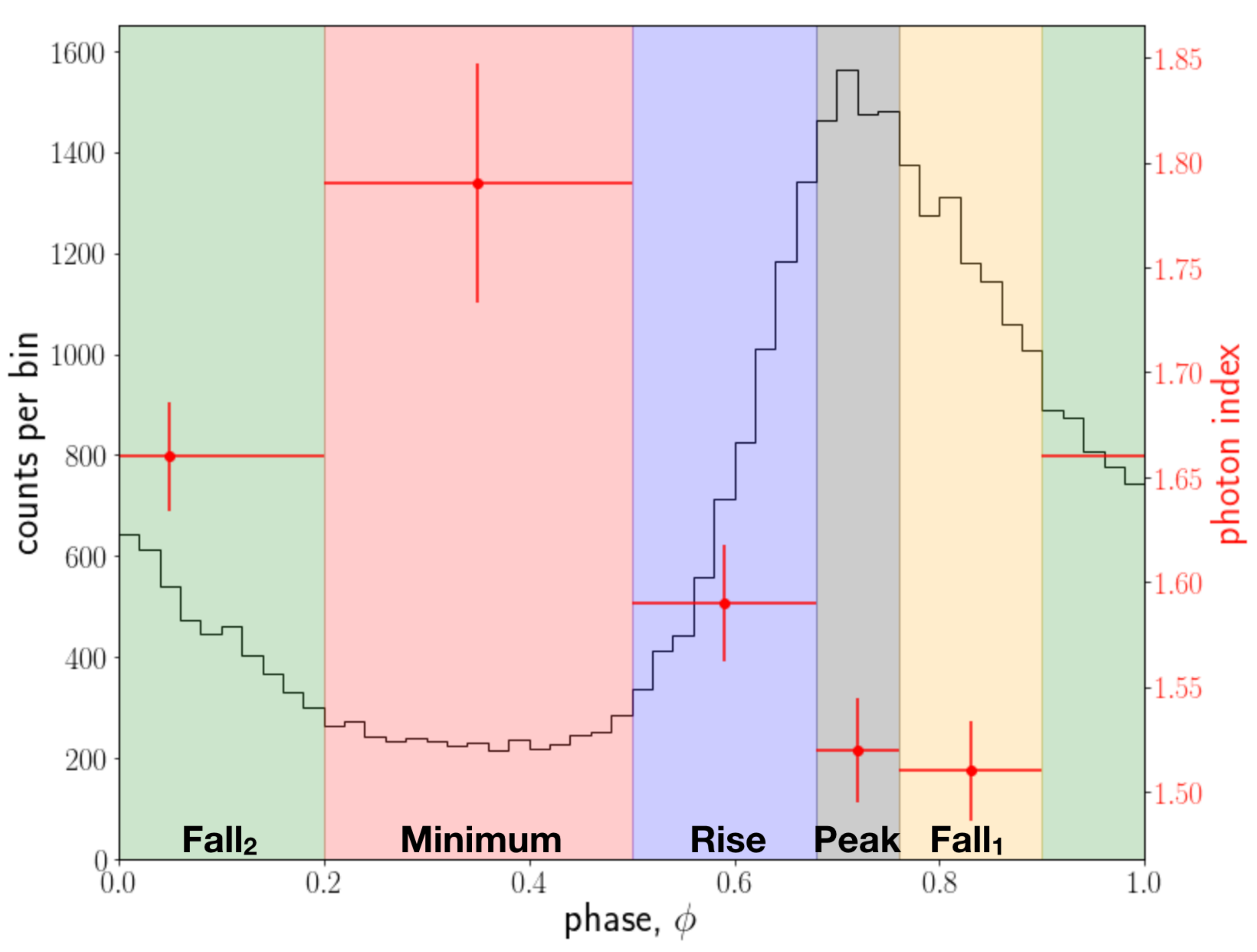}
\caption{Results of absorbed power-law fits to the phase-resolved spectra. The black histogram shows the pulse profile in the 3--40 keV energy range, while the red points and error bars show the photon index, its uncertainties, and the size of the phase bin used for extracting the spectra. The phase bins are pulse minimum (red), rise (blue), maximum (gray), fall$_{1}$ (orange), fall$_{2}$ (green), see Table \ref{tab:spec_tab_pl} for the exact phase definitions for each bin.
The epoch of zero phase is MJD 58239.13162703 TDB, 
the frequency and its derivative are given by Equation (\ref{ephem_sol}).
\label{fig:pr_spec}
}
\end{figure}

Given that strong pulsations were detected by {\sl NuSTAR}, we also explore the phase-resolved spectra of J1617. To do this, we created a phase-folded light curve and defined five phase bins (see Figure \ref{fig:pr_spec} and  Table \ref{tab:spec_tab_pl}). These phase bins were chosen as they belong to similar features observed in the phase-folded light curve,
while also containing enough counts to sufficiently constrain the spectrum (i.e., all spectra have similar uncertainties in the measured photon index except for the minimum phase bin). For example, we chose to include phase bins encapsulating the full-width at 90\% maximum for the peak, while choosing the minimum such that the spectra contained $>1500$ total counts in each detector. 

Prior to fitting the phase-resolved spectra, we froze the $N_{\rm H}$ to the best-fit value of $6.5\times 10^{22}$ cm$^{-2}$
found from the phase-integrated fits. This allowed us to better constrain the photon index as a function of phase. Figure \ref{fig:pr_spec} and Table \ref{tab:spec_tab_pl} show 
how the photon index and the energy flux vary with pulse phase. We find that the photon index 
is larger at pulse minimum and smaller at pulse maximum.

\begin{deluxetable}{lcccc}
\tablecolumns{5}
\tablecaption{
Absorbed power-law fits to the phase-integrated and phase-resolved spectra of J1617. 
}
\label{tab:spec_tab_pl}
\tablewidth{0pt}
\tablehead{
\colhead{Phase} &
\colhead{Range} & 
\colhead{$\Gamma$}  & 
\colhead{$N_{\rm  H}$\tablenotemark{a}} & \colhead{$F_{\rm 3-79\, keV}$\tablenotemark{b}} 
}
\startdata
Integrated & $0<\phi<1$ & 1.59(2) & 6.5(7) & 16.1(3) \\
\tableline 
 Minimum & $0.20<\phi<0.50$ & 1.79(6)  & 6.5\tablenotemark{c} & 4.2(3)\\
 Rise & $0.50<\phi<0.68$ & 1.59(3) & 6.5\tablenotemark{c} & 18.3(6) \\
 Maximum & $0.68<\phi<0.76$ & 1.52(3) & 6.5\tablenotemark{c} & 38(1)\\
 Fall$_{1}$ & $0.76<\phi<0.90$ & 1.51(2)  & 6.5\tablenotemark{c} & 30.4(9)\\
 Fall$_{2}$ & $0.90<\phi<0.20$  & 1.66(3) & 6.5\tablenotemark{c} & 13.1(4)\\
\enddata
\tablenotetext{a}{Absorption in units  of  $10^{22}$ cm$^{-2}$.}
\tablenotetext{b}{Unabsorbed energy flux in units of $10^{-12}$ erg cm$^{-2}$ s$^{-1}$.}
\tablenotetext{c}{$N_{\rm  H}$ frozen to best-fit phase-integrated value.}
\end{deluxetable}

\section{Discussion}
\label{discuss}

\subsection{Implications from the timing results}
\label{timing_results_sec}
The large number of counts 
and long continuous observation (with the exception of gaps due to Earth occultation), 
have allowed us to measure J1617's frequency and its derivative 
with high precision. The advantages of a binning-free timing analysis have also been demonstrated. The hard X-ray pulse profile of J1617 can be described using only three Fourier harmonics, despite the fact that it is non-thermal emission.   
The hard X-ray pulsed fraction is also found to be large and to increase with energy, regardless of the pulsed fraction definition used.

\begin{figure}
\includegraphics[trim={0 0 0 0},scale=0.30]{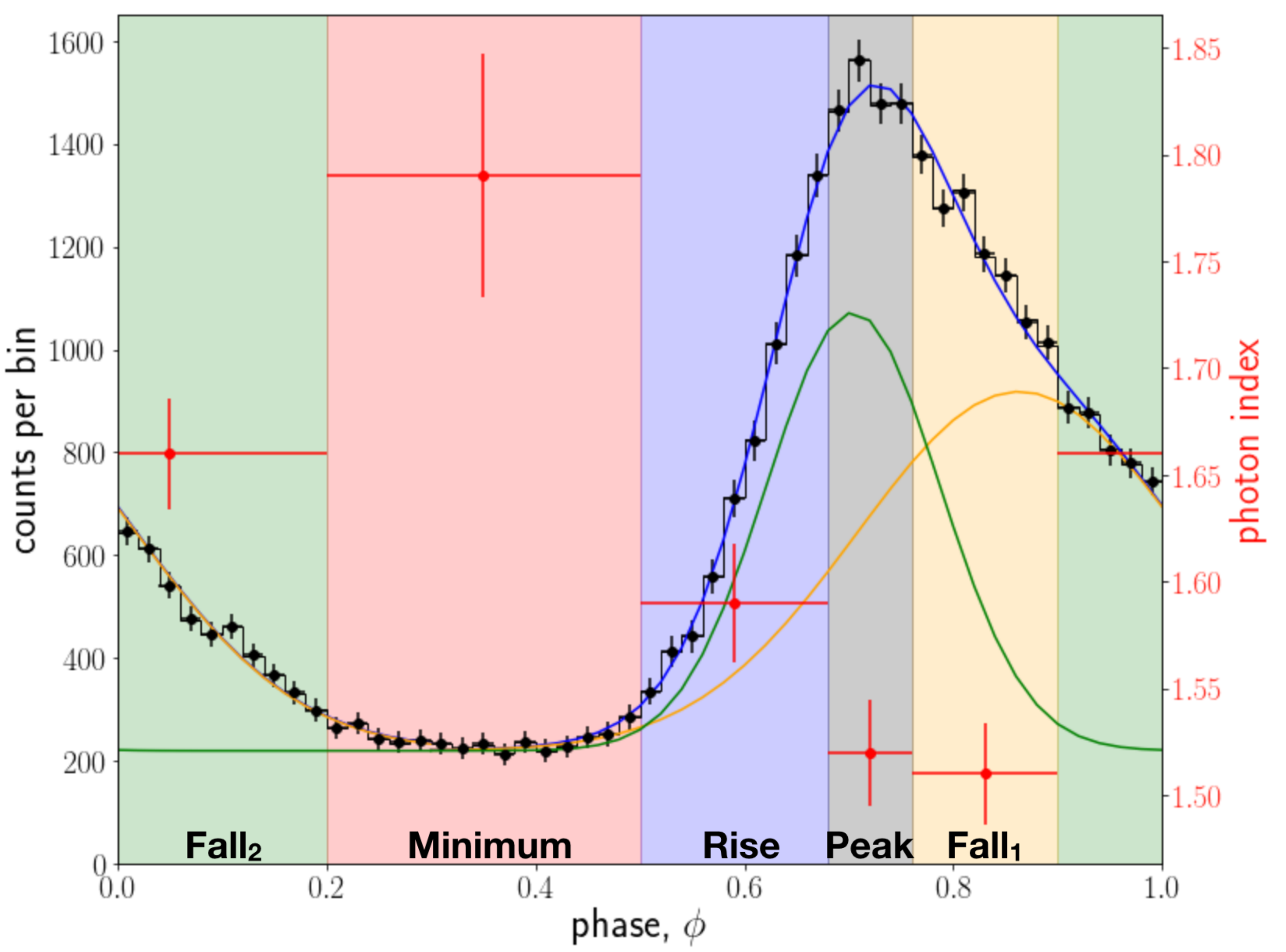}
\caption{The same as Figure \ref{fig:pr_spec} (now with black points and errorbars over plotted on the histogram).
The blue line shows the two-Gaussian fit, 
with the Gaussian components shown as the orange and green lines.
\label{fig:pr_gauss}
}
\end{figure}

The NuSTAR pulse profile is similar to what has been observed at slightly lower X-ray energies
(i.e., down to $\sim2$ keV) by XMM-Newton and RXTE PCA \citep{2002nsps.conf...64B,2015MNRAS.449.3827K}, showing an asymmetric pulse profile with a rise that is faster than the decay. J1617's X-ray pulse is found to be much broader than the radio pulse, having an X-ray duty cycle of about 0.7  compared to $\sim0.08$ at radio wavelengths. Previous studies of several other pulsars exhibiting broad X-ray pulses (e.g., PSR B1509--59, PSR B0540--69) have pointed out that these broad pulses can be fit by two Gaussians, suggesting that the single broad pulse can be decomposed into two individual but overlapping pulses (see e.g. \citealt{1999A&A...351..119K,2001A&A...375..397C,2012ApJS..199...32G}). We also find that J1617's pulse can be well fit by a two-Gaussian model with maximums separated by $0.16$ in phase and having widths of 0.08 and 0.16 in phase (see Figure \ref{fig:pr_gauss}). Furthermore, this
model fits the binned light curve  better than the three-harmonic model 
across a wide range of different bin sizes, which, given better statistics, could have some implications for the interpretation of the phase resolved spectroscopy results (see Section \ref{phase_res_spec_dis}).

\subsection{Phase-Integrated Spectra}
\label{phase_int_spec_sec}
\begin{figure}
\includegraphics[trim={0 0 0 0},scale=0.3]{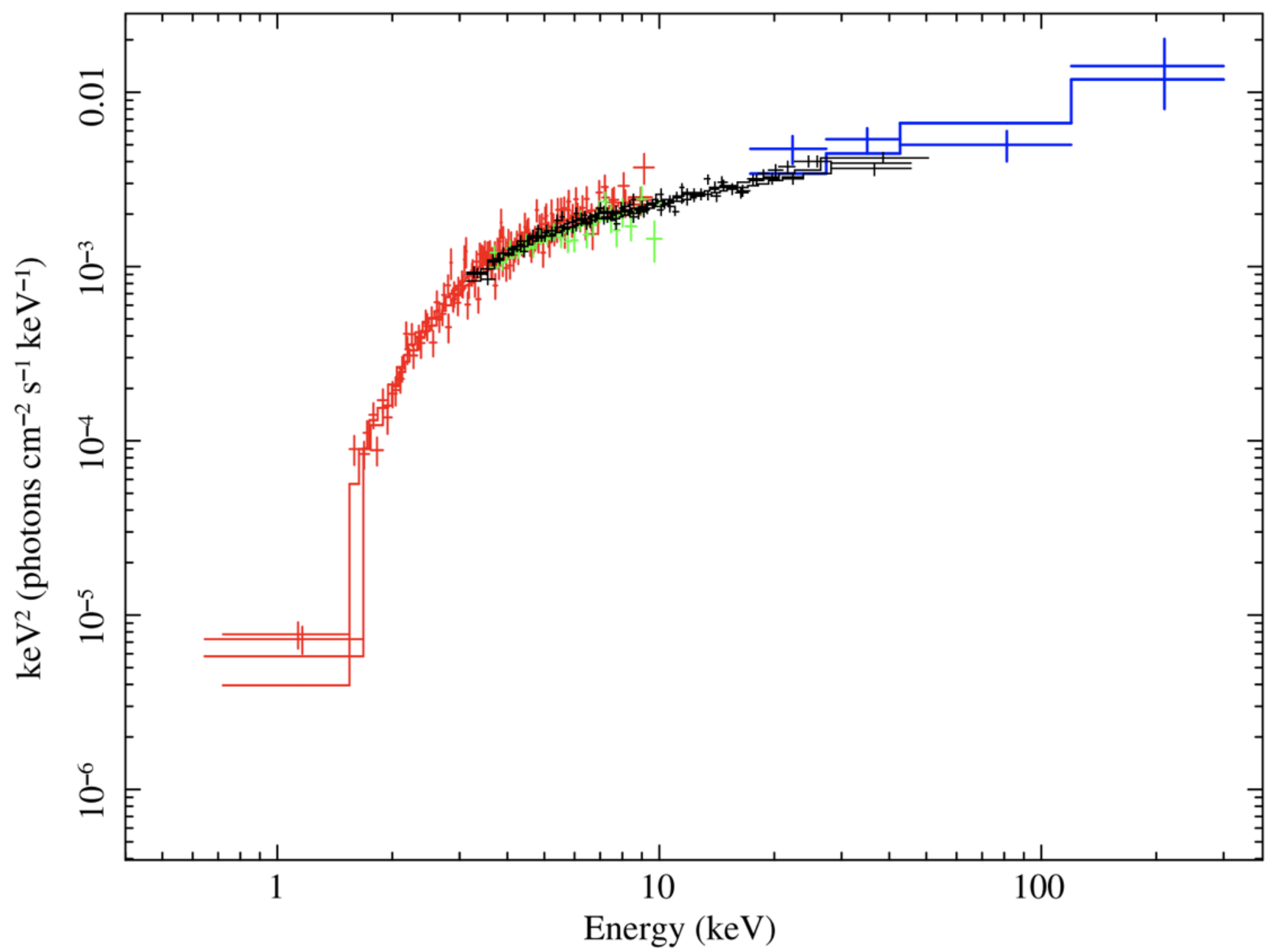}
\caption{X-ray flux spectrum (1-300 keV) 
of J1617.
The XMM-Newton MOS (red), BeppoSAX MECS (green), and INTEGRAL IBIS/ISGRI (blue) data 
and their fits with the absorbed power-law model 
are adopted from \cite{2007MNRAS.380..926L}.
The phase-integrated NuSTAR spectrum and model
(see Table \ref{tab:spec_tab_pl}) are plotted in black.
\label{fig:land_spec}
}
\end{figure}

The best-fit photon index for J1617's phase-integrated spectrum is $\Gamma=1.59(2)$, which is somewhat larger than the $\Gamma\approx1.1-1.4$ found in previous studies at soft X-ray energies\footnote{ We note  that  these older studies use a different absorption model than used here (e.g., {\tt wabs} is used by \citealt{2009ApJ...690..891K}), which leads to the differences in the absorbing column density; however,  this  has a very small effect in the  NuSTAR energy range.} 
(see e.g., \citealt{2002nsps.conf...64B,2007MNRAS.380..926L,2009ApJ...690..891K}). At higher X-ray energies, J1617's 17--300 keV INTEGRAL spectrum is still well fit by a power-law model, but with a  larger photon index $\Gamma\approx1.9$ than in the soft X-ray band, albeit with large  uncertainties $\Delta\Gamma\approx0.3-0.4$ \citep{2007MNRAS.380..926L,2015MNRAS.449.3827K}. 
\cite{2007MNRAS.380..926L} also fit the combined XMM+BeppoSAX+INTEGRAL spectra and found $\Gamma=1.42^{+0.12}_{-0.10}$ (and absorbing column density $N_{\rm H}=3.87^{+0.36}_{-0.28}\times10^{22}$ cm$^{-2}$) which we overplot with the phase-integrated NuSTAR spectrum in Figure \ref{fig:land_spec} for comparison. The difference in the photon indices measured at hard and soft X-ray energies may suggest that there is a  spectral break or some amount of spectral curvature in J1617's X-ray spectrum. We checked this possibility by fitting both an absorbed broken power-law and an absorbed log parabola model to J1617's NuSTAR spectrum. However, neither of these models improved  the  quality of the fit.
Additionally, Figure \ref{fig:land_spec} shows that, qualitatively, the NuSTAR spectra nicely connect the XMM-Newton/BeppoSAX and INTEGRAL spectra.
Lastly, the  flux measured  by NuSTAR is in agreement with  those previously measured in the 2--100  keV energy range \citep{2007MNRAS.380..926L,2015MNRAS.449.3827K}. Therefore, we conclude that the evidence for spectral curvature is currently rather weak.

\subsection{Phase-Resolved Spectra}
\label{phase_res_spec_dis}

The phase-resolved spectra of J1617  show that the photon index changes by $\Delta\Gamma\approx0.3$ between pulse maximum and pulse  minimum. Regarding the pulse minimum, the softness of the PWN spectrum compared to the pulsar spectrum \citep{2009ApJ...690..891K} could potentially lead to the larger measured photon index during the off-pulse as the pulsar is not resolved from the PWN in the NuSTAR image. In other words, if there is no (or little) off-pulse emission from the pulsar itself, then the spectrum could simply be dominated by the PWN at pulse minimum. As mentioned in Section \ref{phase_int_spec}, we expect the PWN to contribute $\sim$ 10\% of the phase-integrated flux from the source. However,  
the relative PWN
contribution should be
much larger when compared to the off-pulse flux. Extrapolating the inner (within $\sim1''$ of the pulsar) and outer (within $\sim1'$ of the pulsar) PWN fluxes, using the  best-fit  models from \cite{2009ApJ...690..891K}, gives a total
(inner + outer) flux of $\sim2\times10^{-12}$ erg cm$^{-2}$ s$^{-1}$ in  the  3--79  keV band.
At pulse minimum, J1617's  flux is only a factor of 2 larger than the anticipated PWN flux, suggesting that the PWN significantly contaminates the off-pulse spectrum. In comparison, the PWN flux is only $\sim5\%$ of the peak flux, while the minimum flux is $\sim10\%$ of the peak flux (or also $\sim5\%$ if the anticipated PWN flux is subtracted).

In an attempt to determine how much the PWN spectrum can impact the derived photon indices, we have also carried out fits of the phase-resolved spectra including the expected contributions from the PWN. To do this, we included two additional absorbed power-law models into our original fits (i.e., for the inner and outer PWN), with their photon indices and fluxes frozen to the values from \cite{2009ApJ...690..891K}. After refitting this new model, all photon indices remained within their uncertainties except for the `minimum' phase interval. The minimum phase photon index did not significantly change after accounting for the PWN emission, $\Gamma=1.91(9)$ versus 1.79(6).   Given this statistically insignificant difference, we conclude that the PWN is unlikely to be responsible for the measured spectral variability as a function of pulsar phase.

On the other hand, the photon index minimum is consistent with occurring near pulse maximum, regardless of whether or not the pulse can be best described as either a single broad pulse, or by two overlapping pulses (see Figure \ref{fig:pr_gauss}). In the single pulse scenario, the photon index maintains its minimum across the single broad peak before becoming larger during the off-pulse emission. Alternatively, in the two pulse scenario, one might expect a change in photon index at or between the two sub-pulse peaks, as has been observed in several other pulsars (see e.g., \citealt{2012ApJS..199...32G}). Unfortunately, we lack the counts to divide the pulse into more phase bins for our phase-resolved spectral fits, thus cannot determine if there is any difference in the photon-index coincident with the two sub peaks (or their overlap) in this scenario.

\subsection{Comparison with other young and energetic pulsars}

There are a number ($\sim20$) of young, energetic rotation-powered pulsars whose non-thermal pulsed emission has been detected above $\sim20$ keV (see \citealt{2015MNRAS.449.3827K} for a review). These pulsars have been further subdivided into GeV-loud and GeV-quiet categories (see, e.g., \citealt{2014MNRAS.445..604W,2015MNRAS.449.3827K}). The GeV-loud pulsars exhibit $\gamma$-ray emission with 
spectral energy distributions (SEDs) peaking at a few GeV. Additionally, $\sim80\%$ of these pulsars show two or more peaks in their GeV and hard X-ray pulse profiles
(e.g., Crab, Vela). On the other hand, the GeV-quiet pulsars (sometimes called MeV pulsars because their SEDs are expected to peak at MeV energies; e.g., \citealt{2017ifs..confE...6H}) also share several similar attributes, including 
under-luminous (or no) pulsed GeV emission, a single peak radio pulse, 
and a single broad pulse detected in the hard X-ray band ($\sim 10$--100 keV; \citealt{2014MNRAS.445..604W}). They also often show a large offset in phase between X-ray and radio pulse peaks. It is also notable that $\approx70\%$ of these $\sim20$ pulsars have associated TeV PWNe \citep{2015MNRAS.449.3827K}.

J1617 can be considered to be an MeV pulsar for several reasons. The first is that, as demonstrated above, its hard X-ray pulse profile is single-peaked and broad. Also, no pulsed GeV emission has been detected from J1617\footnote{It  should be  noted, however, that there is  an extended  GeV source coincident with J1617 \citep{2020ApJS..247...33A}, possibly its relic PWN.} \citep{2013ApJS..208...17A}, and several models predict its high energy SED likely peaks at MeV energies, which is also consistent with the upper limits derived from the Fermi-LAT data (see, e.g., \citealt{2014MNRAS.445..604W,2018NatAs...2..247T}). Unfortunately, there were no radio observations contemporaneous with our NuSTAR observation, so we are unable to measure potential offsets between the radio and hard X-ray pulse peaks.

In the NuSTAR band, J1617 shows similar properties to the young and energetic PSR B1509--58 (B1509 hereafter), which is the archetypal MeV pulsar.  For instance, these two pulsars both have single broad X-ray pulses and show photon indices that are larger at phase minimum and smaller at phase maximum \citep{2016ApJ...817...93C}. In contrast,  the GeV-loud Crab pulsar \citep{2010ApJ...708.1254A} shows two narrower peaks separated by the so-called ``bridge'' in its hard X-ray pulse profile. Its hard X-ray spectrum has a larger photon index at the two pulse peaks and a smaller photon index in the bridge region \citep{2015ApJ...801...66M}.  These differences have led some authors to suggest that the single broad pulse of B1509 is actually composed of two overlapping pulses (see e.g., \citealt{1999A&A...351..119K,2001A&A...375..397C,2012ApJS..199...32G}). As mentioned previously, we also find that J1617's hard X-ray pulse can be well described by two overlapping Gaussian pulses. B1509 is much brighter than J1617, so its photon index was accurately measured by NuSTAR across 26 phase bins, compared to the five bins used here. The better statistics provided by B1509 shows that the smallest photon index is actually offset by about $\sim0.06$ in phase from the maximum phase and occurs in the phases where the two pulses heavily overlap \citep{2016ApJ...817...93C}. Unfortunately, due to the faintness of J1617, we lack the statistics to measure the photon index in small enough phase bins to see where the minimum photon index occurs within the pulse.

One similarity that B1509 and the Crab have are that their off-pulse emission is compatible with having the largest photon indices \citep{2015ApJ...801...66M,2016ApJ...817...93C}. However, it is also important to point out that both the Crab's and B1509's hard X-ray off-pulse emission is heavily contaminated by their bright PWNe. This is less so for J1617, but we still find a large photon index ($\Gamma\approx 1.9$) during the off-pulse emission, even after accounting for the expected PWN contribution. Future hard X-ray observations of J1617, contemporaneous with radio timing observations, would allow one to compare to measure the phase offset between the X-ray and radio peaks. Additionally, future sensitive MeV observatories such as AMEGO or e-ASTROGAM (\citealt{2019BAAS...51g.245M,2018JHEAp..19....1D}) may allow for the detection of J1617 as well as a measurement of its MeV spectrum to constrain the peak emission energy. Such observations could confirm the MeV pulsar nature of J1617.

\acknowledgements
We are grateful to Bettina Posselt for her help in preparing the {\sl NuSTAR} proposal and Karl Forster for his help with planning this observation.  We thank the anonymous referee for carefully reading the manuscript and providing useful comments. Support for this work was provided by the National Aeronautics and Space Administration through
the NuSTAR award 80NSSC17K0640. JH acknowledges support from an appointment to the NASA Postdoctoral
Program at the Goddard Space Flight Center, administered by the USRA through a contract with NASA. 

\appendix

\section{Pulse profile as a sum of Fourier harmonics}

A nearly periodic signal 
${\cal F}(\phi)$ 
(whose period may vary slowly), normalized in such a way that $\int_0^1 {\cal F}(\phi)\, d\phi =1$,
can be presented as the sum of Fourier harmonics:
\begin{equation}
{\cal F}(\phi) = 1 + \sum_{k=1}^K (a_k \cos2\pi k\phi + b_k\sin2\pi k\phi) = 1 + \sum_{k=1}^K s_k \cos2\pi  (k\phi-\psi_k)\,,
\label{eq:fourier}
\end{equation}
where 
\begin{equation}
\phi=\phi(t) =\phi_0 + f (t-t_0) + \dot{f} (t-t_0)^2 /2 + \ldots 
\label{eq:phase}
\end{equation}
is the phase, $f$ and $\dot{f}$ are the frequency and its time derivative, $t_0$ is the reference time, 
\begin{equation}
    a_k=s_k\cos2\pi \psi_k \quad {\rm and} \quad b_k=s_k\sin2\pi  \psi_k
\label{eq:fourier_ceff}
\end{equation}
are the Fourier coefficients for $k$-th harmonic,
which determine the harmonic's amplitude $s_k$, power $s_k^2$, and phase $\psi_k$:
\begin{equation}
    s_k^2=a_k^2+b_k^2,\quad \quad
    \tan2\pi  \psi_k=b_k/a_k\,.
    \label{eq:power_phase}
\end{equation}

If 
the observed signal is a sequence of stochastic (e.g., Poisson-distributed) discrete events (e.g., photon detections), numbered as $i=1, 2, \ldots, N$, with event phases $\phi_i=\phi(t_i)$, then the expected values and variances of 
the Fourier coefficients 
can be calculated as follows:
\begin{equation}
    a_k =
    2\langle \cos2\pi k\phi\rangle =\frac{2}{N}\sum_{i=1}^N \cos2\pi k\phi_i, \quad\quad
    b_k  = 2\langle\sin2\pi k\phi\rangle = \frac{2}{N}\sum_{i=1}^N \sin2\pi k\phi_i.
\label{eq:ak_bk}
\end{equation}
and
\begin{equation}
\sigma^2_{a_k} = 
\frac{4}{N}\left(\langle\cos^2 2\pi k\phi\rangle 
- \langle\cos 2\pi k\phi\rangle^2\right) = 
\frac{1}{N}\left(2+a_{2k} -  a_k^2\right)\,,
\label{eq:sigma_ak}
\end{equation}
\begin{equation}
\sigma^2_{b_k} = 
\frac{4}{N}\left(\langle\sin^2 2\pi k\phi\rangle 
- \langle\sin 2\pi k\phi\rangle^2\right) = 
\frac{1}{N}\left(2-a_{2k} -   b_k^2\right)\,,
\label{eq:sigma_bk}
\end{equation}

Monte-Carlo (MC) simulations, which we use in this work for calculating the
uncertainties of  
$a_k$ and $b_k$, as well as of various functions of these measured Fourier coefficients, have shown a good agreement with Equations (\ref{eq:sigma_ak}) and (\ref{eq:sigma_bk}).

Substituting 
the 
values of $a_k, b_k$ (or $s_k, \psi_k$) in Equation (\ref{eq:fourier}), we obtain a
folded periodic light curve (pulse profile) ${\cal F}(\phi)$, for a given frequency and its derivative.
The representation of a pulse profile as a sum of Fourier harmonics is particularly convenient when the main contribution to the signal comes from a small 
number of harmonics $K$ (e.g., $K=3$ for J1617). 

The profile normalization can be, of course, changed if needed. For instance, ${\cal F}(\phi)$ should be multiplied by $N$ if we want the area under the profile to be equal to the number of events.  The uncertainty of the profile is determined by the uncertainties (standard deviations) of the Fourier coefficients, $\sigma_{a_k}$ and $\sigma_{b_k}$, and uncertainties of $f$ and $\dot{f}$.
An example of this profile for J1617's pulsations is shown in Figure \ref{fig:all_ampls_corr}.

\section{Searching for pulsations, and measuring 
frequency and its derivative} 

The Fourier representation 
is also useful for pulsation searches  and the measurements of  period (frequency) and its derivatives.
In high energy astronomy the $Z_K^2$ statistic (e.g., \citealt{1983A&A...128..245B}),
\begin{equation}
    Z_K^2=
    Z_K^2(f,\dot{f}; t_0)= \frac{N}{2} \sum_{k=1}^K s_k^2 = \frac{N}{2} {\cal S}_K^2,
\label{eq:Zn-square}
\end{equation}
is 
often used for these goals (here and below we assume that higher than first order frequency derivatives can be neglected).
The maximum number $K$ of harmonics  
can be estimated from the so-called H-test \citep{1989A&A...221..180D}: $K$ is the number of harmonics that maximizes the quantity $Z_k^2 -4k +4$ over a 
set 
of $k$ values, $1\leq k \leq k_{\rm max}$, and the H-statistic is defined as $H=Z_K^2 - 4K +4$.

If the pulsation frequency is approximately known (like in the case of J1617), one can pinpoint its actual value (and the value of frequency derivative, if feasible) by locating the frequency and its derivative at which $Z_K^2$ reaches a maximum over a relatively small grid of trial $f$, $\dot{f}$ values. If the frequency is not known even approximately, one has to analyze maxima of $Z_K^2$ (or $H$) over a much broader grid in order to detect pulsations. The maximum, $Z_{K,\rm max}^2$, corresponding to true pulsations  should be high enough to distinguish it from the Poissonian noise, for which $Z_K^2$ is distributed as $\chi_{2K}^2$. For instance, for $K=1$, the probability that
$Z_{1}^2$ in the noise power spectrum exceeds the found peak $Z_{1, \rm max}^2$ (i.e., the probability that the peak is generated by noise) 
can be estimated as
\begin{equation}
    {\rm Prob}(Z_{1}^2 > Z_{1,\rm max}^2) = 1 - \left[1-\exp\left(-Z_{1, \rm max}^2/2\right)\right]^{N_{\rm tr}} \approx N_{\rm tr} \exp(-Z_{1,\rm max}^2/2)\,,
\end{equation}
where $N_{\rm tr}$ is the number of statistically independent trials (e.g., $N_{\rm tr}= (f_{\rm max}-f_{\rm min}) T_{\rm span}$ if only the frequency is varied, in the range from $f_{\rm min} < f < f_{\rm max}$), and the approximate equality implies $N_{\rm tr} \exp(-Z_{1, \rm max}^2/2) \ll 1$.   

The uncertainties of the 
found $f,\dot{f}$ values depend on the observation time span and the height of $Z_K^2$ maximum.
In the case of purely sinusoidal pulsations ($K=1$) observed without gaps and with time resolution 
much shorter than the pulsation period, 
the uncertainties at the 68\% confidence level 
can be estimated as
\begin{equation}
 \sigma_f = (\sqrt{3}/\pi)\, T_{\rm span}^{-1} \left(Z_{\rm 1,max}^2\right)^{-1/2},\quad\quad  
 \sigma_{\dot{f}}= (6\sqrt{5}/\pi) T_{\rm span}^{-2} \left(Z_{1,\rm max}^2\right)^{-1/2}
    \label{eq:deltaffdot}   
\end{equation}
(similar estimates, in different notations, can be found in 
\citealt{2002AJ....124.1788R}
for the Fourier power peak at the fundamental frequency; the estimate for $\sigma_f$ was independently derived by \citealt{2012ApJ...744...81C}).
If we neglect harmonics higher than $k=1$ in the J1617 power spectrum and do not take the observation gaps into account, 
Equations (\ref{eq:deltaffdot}) give 
$\sigma_f = 2.0\times 10^{-8}$ Hz, 
$\sigma_{\dot{f}} = 6.4\times 10^{-13}$ Hz s$^{-1}$. 
 
To estimate the uncertainties with allowance for higher harmonics and observation gaps,
we generated 1000 synthetic 
periodic signals with the measured frequency and its derivative for 
 the same time span and time gaps 
as in our data, and the amplitudes and phases of 3 harmonics  
equal to the values found for the 3--79 keV band. We fixed 
the average event rate for each segment between the gaps and varied the number of events and times of arrival according to the Poisson statistic.
For each of the 1000 realizations we found the position of $Z_3^2$ maximum in the $f$, $\dot{f}$ plane and measured the mean and the standard deviation of these positions.
These simulations 
yielded the uncertainties $\sigma_f= 1.8\times 10^{-8}$ Hz, $\sigma_{\dot{f}}=5.6\times 10^{-13}$ Hz s$^{-1}$, virtually coinciding with those given by Equations (\ref{eq:deltaffdot}).

We used similar 
simulations to determine the uncertainties of the Fourier coefficients, pulsed fractions, and pulse profiles for various energy bands, described in Section \ref{X-ray timing sec}. 

\section{Different definitions of pulsed fraction}

Most naturally, the pulsed fraction is defined as the {\em ratio of areas} under the varying part of the light curve to the total area,
\begin{equation}
    p_{\rm area}= 1-{\cal F}_{\rm min}\,,
\label{eq:pf-area}
\end{equation}
where ${\cal F}_{\rm min}$ is the minimum value of ${\cal F}(\phi)$.

One can also define the ``pulsed fraction'' as
\begin{equation}
    p_{\rm amp} = \frac{{\cal F}_{\rm max} - {\cal F}_{\rm min}}{{\cal F}_{\rm max} + {\cal F}_{\rm min}}
    \label{eq:pf_amplitude}
\end{equation}
It is sometimes called ``peak-to-peak'' or ``max-to-min'' pulsed fraction. We believe the 
{\em modulation amplitude} is a more appropriate term for this quantity, so we denote it $p_{\rm amp}$.

Both the `area pulsed fraction' and, particularly, the 
`modulation amplitude' suffer from large uncertainties of measured values of ${\cal F}_{\rm min}$ (and ${\cal F}_{\rm max}$, to a lesser extent), especially in the cases when ${\cal F}_{\rm min}$ is close to zero. To 
avoid this uncertainty, one can characterize pulsations by the {\em root mean square} (rms) of the signal ${\cal F}(\phi)$, which is sometimes called the `rms pulsed fraction' (although it is a misnomer -- $p_{\rm rms}$ is not a pulsed fraction per se):
\begin{equation}
 p_{\rm rms} = \left\{\int_0^1 [{\cal F}(\phi) -1]^2 d\phi \right\}^{1/2} = \left(\frac{1}{2}\sum_{k=1}^K s_k^2\right)^{1/2} = 
\frac{{\cal S}_K}{\sqrt{2}}= \left(\frac{Z_K^2}{N}\right)^{1/2}\,,
\label{eq:pf_rms}
\end{equation}
where the harmonic powers $s_k^2$
and the $Z_K^2$ statistic 
are calculated for the best-fit signal frequency and frequency derivative (i.e., $Z_K^2=Z_{K,\rm max}^2$). Thus, Equation (\ref{eq:pf_rms}) shows that the 
signal rms  can be immediately calculated once $Z_K^2$ is known. 
Evaluation of 
$p_{\rm rms}$ does not involve a measurement of the uncertain ${\cal F}_{\rm min}$, and it is more convenient (and less prone to uncertainties) than $p_{\rm area}$ and $p_{\rm amp}$ if the number of harmonics involved is not too large.

For an illustrative example of a sinusoidal signal 
\begin{equation}
    {\cal F}(\phi) = 1 + q \cos2\pi (\phi-\phi_0)
\end{equation}
we have
\begin{equation}
    a_1 = q \cos 2\pi\phi_0, \quad  b_1 =q \sin 2\pi\phi_0,\quad\quad s_1 =q,\quad \psi_1 = \phi_0,\quad\quad
    Z_1^2 = Nq^2/2
\end{equation}
and
\begin{equation}
p_{\rm area}=p_{\rm amp} =q, \quad\quad
     p_{\rm rms} = \frac{q}{\sqrt{2}}.
     \label{eq:pf_sinus}
\end{equation}
Thus, for sinusoidal pulsations, $p_{\rm rms}$
 is a factor of $\sqrt{2}$ smaller than the two other `pulsed fractions'. For a general case, when the signal contains several harmonics with different phases $\psi_k$, all three `pulsed fractions' are different, and there is no a simple analytical relationship between them.

It should be noted that because $s_k^2$ values in Equation (\ref{eq:pf_rms})
contain not only the signal but also the noise contribution, the $p_{\rm rms}$ estimate is biased toward higher values. 
To correct for this bias, one can subtract the
noise contribution, 
with the aid of Equations (\ref{eq:sigma_ak}) and (\ref{eq:sigma_bk}):
\begin{equation}
\label{p_rms}
    p_{\rm rms} = 
\left[\frac{1}{2} \sum_{k=1}^K 
(a_k^2+b_k^2-\sigma_{a_k}^2 - \sigma_{b_k}^2)\right]^{1/2} =
\left[\left(1+\frac{1}{N}\right) \frac{{\cal S}_K^2}{2} - \frac{2K}{N}\right]^{1/2} \simeq    
    \left(\frac{Z_K^2-2K}{N}\right)^{1/2}.
\end{equation} 
This correction is, however, negligible for strong signals, such as we detect from J1617.

The uncertainty of $p_{\rm rms}$ can be expressed in terms of Fourier coefficients and calculated analytically, unlike $p_{\rm area}$ or $p_{\rm amp}$.
For the case of J1617 we, however, estimated it from MC simulations.

\end{document}